# *Ab Initio* Calculation of Impurity Effects in Copper Oxide Materials

*L.-L. Wang, P. J. Hirschfeld, and H.-P. Cheng*

*Department of Physics, University of Florida, PO Box 118440, Gainesville FL 32611 USA*

## Abstract

We describe a method for calculating, within density functional theory, the electronic structure associated with typical defects which substitute for Cu in the $CuO_2$ planes of high-$T_c$ superconducting materials. The focus is primarily on $Bi_2Sr_2CaCu_2O_8$, the material on which most STM measurements of impurity resonances in the superconducting state have been performed. The magnitudes of the effective potentials found for Zn, Ni and vacancies on the in-plane Cu sites in this host material are remarkably consistent with phenomenological fits of potential scattering models to STM resonance energies. The effective potential ranges are quite short, of order 1 Å with weak long range tails, in contrast to some current models of extended potentials which attempt to fit STM data. For the case of Zn and Cu vacancies, the effective potentials are strongly repulsive, and states on the impurity site near the Fermi level are simply removed. The local density of states (LDOS) just above the impurity is nevertheless found to be a maximum in the case of Zn and a local minimum in case of the vacancy, in agreement with experiment. The Zn and Cu vacancy patterns are explained as due to the long-range tails of the effective impurity potential at the sample surface. The case of Ni is richer due to the Ni atom's strong hybridization with states near the Fermi level; in particular, the short range part of the potential is attractive, and the LDOS is found to vary rapidly with distance from the surface and from the impurity site. We propose that the current controversy surrounding the observed STM patterns can be resolved by properly accounting for the effective impurity potentials and wave-functions near the cuprate surface. Other aspects of the impurity states for all three species are discussed.

# Introduction

Much insight into the superconducting state has traditionally been acquired by studying what destroys it. Shortly after the early work of BCS, it was recognized that while nonmagnetic impurities would not affect thermodynamic properties of a conventional superconductor[1], they would do so in hypothetical non-$s$-wave pairing systems[2]. This sensitivity to ordinary disorder was realized first in the heavy fermion superconductors, whose symmetry class is still under debate, and in the cuprates, where it is now clear that $d$-wave superconducting order is present. Recently, scanning tunneling microscopy (STM) has provided a singular tool to understand the effect of impurities in the superconducting state at the *local* level by measuring the tunneling current which, modulo matrix element effects, is proportional to the local density of states (LDOS) at the tip position. Both in the case of magnetic impurities in conventional superconductors[3], and for nonmagnetic Cu substitutions in the high-$T_c$ cuprates[4-6], impurity bound states have been reported which appear as a bias-dependent pattern of high and low tunneling intensity centered around the putative impurity site (Figure 1, left column). For any tip position near the impurity, a peak structure as a function of bias voltage is observed centered at a resonance energy characteristic of the effective impurity potential (Figure 1, right column).

Local properties of impurities in unconventional superconductors have been reviewed recently by Balatsky *et al*[7]. The existence of impurity bound states of nonmagnetic type in unconventional superconductors was first discussed by Stamp[8], while the use of STM to probe impurity wave functions in a $d$-wave superconductor was proposed by Byers *et al*[9]. Balatsky *et al*.[10] then calculated the bound state wave function in the $d$-wave case appropriate to cuprate superconductors. All these calculations modeled the impurity as a simple point-like potential scatter. Partial confirmation of these predictions was provided by STM, which discovered localized impurity resonances first around apparent "native defects" in the $CuO_2$ plane[4], and then around Zn[5] and Ni[11] substituted for Cu. A different class of apparently ubiquitous "native" impurity resonances in nominally pure samples was identified in Ref. [6] and suggested to correspond to Cu vacancies in the $CuO_2$ plane. While measured resonance energies on native defects and Zn are roughly consistent with fits to bulk transport measurements based on naive point-like potential scattering models[12], the spatial patterns observed were qualitatively different from those predicted. For example, since the $Zn^{++}$ ion has a filled $d$ shell, one expects an $S=0$ state with effective potential of several eV, such that $UN_0 \gg 1$, where $N_0$ is the density of states at the Fermi level. The resonance should therefore occur, in a particle-hole symmetric situation, at small negative energies [10]

$$\Omega \simeq -\frac{\Delta_0}{N_0 U}\frac{1}{\log(8N_0 U)}, \tag{1}$$

where $\Delta_0$ is the gap maximum; and indeed, a sharp peak is observed in STM at about -1.5 meV[5]. On the other hand, in such a picture one expects small current at the impurity site for negative tip bias due to the localized electron on the Zn, but enhanced current at nearest neighbor sites[10]. Empirically, the opposite is seen[5], which has prompted a number of theoretical explanations. These fall into three categories: a) attempts to model the impurity atoms with extended effective potentials within the $CuO_2$ plane (e.g., postulating a significant negative potential at the Zn site but positive on the nearest neighbor site can qualitatively reproduce the experimental pattern for Zn)[13]; b) assuming that tunneling matrix elements between states of interest in the Cu-$O_2$ plane and the exposed Bi-O layer on the cleaved $Bi_2Sr_2CaCu_2O_8$ (BSCCO) surface "filter" the intrinsic LDOS pattern in the $CuO_2$ plane[14,15]; and c) arguments that electronic correlations in the host induce non-local spin exchanges around the impurity site which dominate the scattering[16]. At the same time, it should be noted that native defects and Ni impurities produce spatial patterns and dispersions with STM bias which are qualitatively different from Zn and which are not completely understood.

In this work, we assume that strong local correlations do *not* constitute the most important factor influencing impurity states, and instead ask if a consistent description of some impurity experiments can be obtained by describing the impurity and host wave functions much more accurately than heretofore attempted. The initial strong assumption allows us to use the tools of density functional theory (DFT)[17, 18] to ensure that the full electronic structure of BSCCO and the local chemistry of the impurities are properly accounted for. Although we believe that this is a reasonable attempt to capture some of the qualitative systematics of the impurity problem, and some similar approaches have been tried[19], it is not *a priori* obvious that such a calculation method can succeed. This is because density functional theory is known to fail for the high-$T_c$ parent compounds, which are anti-ferromagnetic Mott insulators. It is believed that a stoichiometric crystal of the BSCCO material itself would be insulating if one could be made; in fact, current BSCCO crystals are probably doped by naturally occurring excess oxygen and Ca, and also frequently stabilized by significant non-stoichiometric quantities of Sr and Bi[20]. Nevertheless, local density approximation (LDA) calculations[21] are known to produce electronic bands which are not too different from those measured by angle-resolved photoemission (ARPES) on the actual as-grown crystals[22]. Furthermore, we anticipate that the magnitude and form of the effective impurity potential will be determined by high-energy states which will not be so sensitive to the correlation effects as the strongly renormalized states near the Fermi level. Ultimately whether or not our approach will be applicable must be judged by its success in accounting for experimentally observed properties.

We are in this work primarily concerned with the physics of impurity wave functions in the *normal state* of the BSCCO material with zero superconducting order parameter, but try to draw conclusions relevant to the superconducting state. We know *a priori* that STM LDOS patterns will be less striking in the normal state, because the resonant behavior in the superconducting state is caused by the removal of states from the vicinity of the Fermi level by pairing. The calculation of superconducting state spectra suitable for direct comparison to experiment is greatly to be desired, but technically beyond the scope of this work. This is because such a calculation requires either a microscopic model of the superconducting pairing interaction, which cannot be calculated from DFT, or an *ansatz* for the off-diagonal self-energy of the system. The first approach is clearly unjustified in view of our ignorance of the underlying microscopic theory of pairing in these systems, while the second is plausible, but requires much more difficult calculations to obtain the necessary energy resolution. We try here instead to extract from normal state results a sense of which physical effects might be relevant to superconducting state experiments, and to calculate effective potentials for different impurities suitable for use in phenomenological models of *d*-wave superconductivity. This is a reasonable way to proceed because impurity energies, of order eV, are much larger than typical pairing energies, of order 40 meV or so, and may therefore be reliably assumed to be unaffected by the onset of pair correlations.

We begin below by discussing the technical details of the calculation of impurity effects using DFT. We then present our results on the structure relaxation of the BSCCO surfaces doped with the three most common impurities and the electronic structures. Next, we address the STM problem and discuss how the states just above the BiO surface layer change in response to the presence of an impurity in the $CuO_2$ plane. Finally, we present our results for effective impurity potentials for the different defects and compare with other calculations and theoretical *ansätze* in the literature.

## Computational Details

In this work, DFT calculations have been used to determine all structural, energetic and electronic results. The Kohn-Sham equations are solved self-consistently in a plane-wave basis set, in conjunction with Vanderbilt ultrasoft pseudopotentials[23, 24], which describe the electron-ion interaction, as implemented in the Vienna *ab initio* simulation program (VASP)[25-27]. Exchange and correlation are described by LDA. We use the exchange-correlation functional determined by Ceperly and Alder[28] and parameterized by Perdew and Zunger[29].

According to experiments[30-33], the BSCCO-2212 crystal has a pseudo-tetragonal substructure unit cell, as seen in Figure 2(a), with dimension of $a \times a \times c$, where $a = 3.814$ Å and $c = 30.52$ Å[32]. The structure can be described as two neighboring slabs of $c/2$ thick along the [001] direction, which are shifted by $a/2$ and weakly bonded by the long Bi-O bonds (3.0 Å). Inside each slab, there is one formula unit of $Bi_2Sr_2Ca_1Cu_2O_8$ (referred as BSCCO-2212). It consists of one Ca layer sandwiched by two $CuO_2$, two SrO, and two BiO layers. In each layer, the bonds formed between metal and oxygen atoms are almost perpendicular to the $c$ axis, except for a strong buckling of the Sr-O bond. The crystal is also slightly distorted to an orthorhombic unit cell with dimension of $\sqrt{2}a \times 5\sqrt{2}a \times c$ in the high $T_c$ phase[31, 33]. Since we are interested in the impurity effect in the normal phase of BSCCO in this study, the tetragonal substructure unit cell is used as the primitive cell.

Due to the weak bonding between two neighboring BiO layers, the BSCCO-2212 surface is nearly always exposed at the cleaved BiO layer. For the calculation of the BSCCO surface doped with impurities, we use surface unit cells (or super cell) with dimension of $\left(2\sqrt{2}a \times 2\sqrt{2}a\right)R45º$ that contain 8 primitive unit cells in the *xy* plane. There are three different Cu atom sites between two impurities in adjacent super cell, which allow us to study the effect of the impurity on the first through third nearest neighbor Cu atoms. In the *z* direction, we use only half of the primitive unit cell $(c/2)$, because the interaction with the other half is so weak that the omission of it does not affect the surface properties in our test calculations. The impurity substitutes one Cu atom in the first $CuO_2$ plane from the top as seen in Figure 2 (b), (c) and (d). The largest calculation consists of 120 atoms. The spacing of the vacuum between the neighboring slabs is about 15 Å.

In all the calculations, the kinetic energy cutoff is 400 eV, or 29.4 Rydberg. The first-order Methfessel-Paxton[34] smearing of 0.2 eV is used. The *k*-point mesh[35] used for the bulk calculation is (10x10x1) in the first Brillouin zone, which gives 15 nonequivalent *k*-points in the irreducible part of the first Brillouin zone because of the body-centered tetragonal symmetry of *I4/mmm*. In the surface calculation, the *k*-point mesh in use is (4×4×1), which gives 3 nonequivalent *k*-points. Convergence tests have been performed with respect to the *k*-point mesh and vacuum spacing. The total energy converges to 1 meV/atom. The ionic structure relaxation is performed with a quasi-Newton minimization using Hellmann-Feynman forces. For ionic structure relaxation, all layers of the slab are allowed to relax until the absolute value of the force on each atom is less than 0.02 eV/Å.

## Results and Discussion

### A. Structure relaxation

For the bulk calculation, we let both the volume and the positions of the atoms inside the tetragonal unit cell relax. The optimized parameters for the unit cell are $a = 3.67$ Å and $c = 30.35$ Å, which are 4 % and 1 % less than the experimental values. These are within the typical error of LDA in calculating the lattice constants of bulk materials. The relative positions of the atoms inside the unit cell change very little from those of the experimental data[32, 33].

The relaxed structures of the BSCCO surfaces are doped with Zn and Ni, the two impurities most commonly substituted for Cu, as well as a Cu vacancy, a promising candidate for the "native defect" resonances seen by STM at low bias. The positions of nearby atoms are shown for these cases in Figure 2 (b), (c) and (d), respectively. The arrow on the atom and the number given indicate the direction and magnitude (Å) of the atom's displacement with respect to its position in the clean crystal in the presence of a free surface at the BiO layer. Most of the structure relaxation is on the impurity sites. The atoms are only relaxed to the near vicinity of the corresponding sites in the clean BSCCO surface. The magnitude of the structure relaxation is not large enough to change any of the bonding order in the layered structure of the BSCCO surface.

It is perhaps worth noting, however, that these atomic displacements around the impurity will lead to significantly renormalized model parameters near the impurity site if an effective low-energy Hamiltonian were to be derived from the uncorrelated wave-functions. For example the local hopping $t$ or exchange $J$ would depend quite sensitively on displacements of these magnitudes.

For a Zn impurity doped BSCCO surface, as seen in Figure 2(b), the Zn atom moves upward by 0.20 Å, the Bi atom above it moves downward by 0.08 Å and the Cu atom below it moves downward by 0.01 Å, etc. This shows that the Zn impurity attracts the Bi atom above it and repels the Cu atom below it, while the Ni impurity repels the Bi above it and attracts the Cu atom below it. The vacancy impurity repels both the Bi and Cu atoms. These different behaviors in the structure relaxation of the BSCCO surface in the presence of different impurities will also be discussed in the section below by evaluating the projected density of states on individual atomic sites.

## B. Density of states

The total density of states (DOS) and partial density of states (PDOS) projected on atomic species are shown in Figure 3. The PDOS is defined as,

$$DOS_\mu(\varepsilon) = \sum_{\mathbf{k}} w_\mathbf{k} \sum_i \delta(\varepsilon - \varepsilon_{i,\mathbf{k}}) |\langle \phi_{\mu,\mathbf{k}}(\mathbf{r}) | \Psi_{i,\mathbf{k}}(\mathbf{r}) \rangle|^2 \qquad (2)$$

where $\phi_{\mu,\mathbf{k}}$ are linear combinations of atomic wave functions that obey the crystalline symmetry and $\Psi_{i,\mathbf{k}}(\mathbf{r})$ are Bloch state wavefunctions, respectively; $w_\mathbf{k}$ is the weight of each $\mathbf{k}$ point, and the indices $\mu$ and $(i,\mathbf{k})$ are labels for atomic orbitals and Bloch states, respectively. The projection provides a useful tool for analyzing the band structure and the density of states. For a clean BSCCO surface, as seen in Figure.3(a), the occupied states from −2 eV to the Fermi level are dominantly from the Cu and O(Cu) atoms in the $CuO_2$ layers, where the element in parentheses labels the layer in which the O atom is located. As for the unoccupied states just above the Fermi level, they are mostly from the Bi, O(Bi) and O(Sr) atoms. Part of the Bi and O(Bi) derived states extend below the Fermi level. They contribute significantly to the states on the Fermi level after the Cu and O(Cu) atoms. The BiO bands and $CuO_2$ bands tend to pin the Fermi level between them as discussed by Krakauer and Pickett[21]. For the BSCCO surfaces doped with impurities, global features of DOS still hold, as seen in Figure 3 (b)-(d). The impurities give rise to several significant changes in the total DOS, however. For the BSCCO surface doped with Zn or vacancy impurities, there is no change near the Fermi level, due in the case of Zn to the fact that the filled Zn $3d$ bands lie at −7 eV. In the case of Ni, there is a significant increase of the states just below the Fermi level due to the high-lying Ni $3d$ bands. Although Ni is nominally a spin-1 impurity, within spin-polarized calculations no spontaneous magnetization for the states around the Ni was found here.

In order to understand the difference in structural relaxation of the BSCCO surfaces doped with different impurities, the PDOS projected on $3d$ orbitals of the Cu atom just below the impurity and $6p$ orbitals of the Bi atom above the impurity are shown in Figure 4(a) and (b), respectively. In each panel, the PDOS from the BSCCO surfaces doped with Zn, Ni and Cu vacancy are compared to the reference PDOS with no impurity. As seen in Figure 4(a), in the presence of a Ni impurity, the density of the anti-bonding Cu $3d$ states is suppressed at higher energies (−2 to −1 eV) and enhanced at lower energies (−3 to −2 eV) compared with the clean surface, whereas in the cases of a Zn and vacancy impurities the anti-bonding Cu $3d$ states in the range from −3 to −2 eV are shifted toward higher energy compared with the clean surface. This is consistent with the Cu atom below the impurity tending to relax away from the O(Sr) atom in the bottom SrO layer and toward the Ni impurity, while it relaxes toward the O(Sr) atom and away from the Zn and vacancy impurities. A similar analysis can also be done for the PDOS on Bi $6p$ states in Figure 4(b). The bonding Bi $6p$ states around −6 eV below the Fermi level are shifted toward lower energy in the presence of a Zn impurity, while they are shifted toward higher energy in the presence of a Ni and vacancy impurities. This means the Bi atom over the Zn impurity binds more strongly to the surface relative to the Ni and vacancy impurities. This is indeed the case, as shown in Figure 2(b)-(d) by the relaxed position of the Bi atoms.

## C. Effective impurity potentials

In order to fully understand the influence of the impurities via the BiO and SrO blocking layers on the LDOS above the surface, we examine the effective impurity potential, $\Delta V_{eff}(\mathbf{r})$, in Figure 5 and Figure 6. In DFT, the total energy of a many electron system can be written as,

$$E_{elec} = T_e + E_{eI} + E_{ee} = T_e + U_e \quad (3)$$

where,

$$T_e = \sum_{j,\sigma} f_{j\sigma} \langle j\sigma | T_\sigma | j\sigma \rangle, \quad (3a)$$

$$E_{eI} = \sum_{j,\sigma} f_{j\sigma} \langle j\sigma | V_{I,\sigma} | j\sigma \rangle, \quad (3b)$$

$$E_{ee} = \int d\mathbf{r}^3 \left( \rho_+(\mathbf{r}) + \rho_-(\mathbf{r}) \right) \left( \varepsilon_H + \varepsilon_{xc} \right), \quad (3c)$$

$$\rho_\sigma(\mathbf{r}) \quad \rho_\sigma(\mathbf{r})\, \varepsilon_H = \int \frac{1}{2} d\mathbf{r}'^3 \frac{\left( \rho_+(\mathbf{r}') + \rho_-(\mathbf{r}') \right)}{|\mathbf{r} - \mathbf{r}'|}. \quad (3d)$$

The effective potential is the functional derivative of the energy,

$$V_{eff,\sigma}(\mathbf{r}) = \frac{\delta U}{\delta \rho_\sigma(\mathbf{r})} \quad (4)$$

In equation 3-4, σ is the spin index (+ or −), $T_\sigma$ and $V_{I,\sigma}$ are operators of kinetic energy and potential energy between nuclei and electron interaction, $|j,\sigma\rangle$ indicate $j^{th}$ eigenstate of the system (e.g., a Bloch state or a molecular orbital), $f_{j,\sigma}$ is the Fermi distribution function, $\rho_\sigma(\mathbf{r})$ is the spin charge density distribution, $T_e$ is the kinetic energy of a non-interacting electron system, $\varepsilon_H$ is the "Hartree" energy functional, and $\varepsilon_{xc}$ is the exchange-correlation energy functional. They are both functionals of $\rho_\sigma(\mathbf{r})$. In the local spin density approximation, $\varepsilon_{xc}$ depends only on the densities, and is the exchange-correlation energy per electron of a uniform electron gas with charge density $\rho_\sigma(\mathbf{r})$. The quantity $\Delta V_{eff}(\mathbf{r})$ is obtained by subtracting the effective potential of the clean surface defined within DFT from that of the impurity doped surface. After examining the effective potential with and without the structure relaxation, we find the effect of the structure relaxation on the magnitude of the effective impurity potential is very small on the impurity site and on the plane 1.5 Å above the surface where the STM images are taken. In order to reduce the numerical noise, we therefore use the unrelaxed structure of the surface to calculate these quantities.

Figure 5 shows the iso-energetic surfaces of $\Delta V_{eff}(\mathbf{r})$ at two different magnitudes for BSCCO-2212 cleaved at the BiO surface and doped with different impurities in the first $CuO_2$ plane. The red (darker) and yellow (lighter) indicate that the effective impurity potential is negative and positive, respectively, in that region. Figure 5(a) shows that the effective impurity potential is large and positive in the near vicinity of the Zn impurity inside a roughly spherical region. Within roughly 1 Å from the Zn impurity, the effective impurity potential becomes negative, however. The negative iso-energetic surface spans a cylindrical region, whose axis is along the [001] direction. This shows that the influence of the Zn impurity can easily extend beyond the $CuO_2$ layer in the [001] direction toward the BiO layer and the effective impurity potential oscillates as the distance from the impurity increases. These features can also be seen in Figure 5(b) from the iso-energetic surfaces of $\Delta V_{eff}(\mathbf{r})$ at a smaller magnitude. On the $CuO_2$ layer, the effective impurity potential oscillates on the neighboring sites. Near the BiO layer, the effective impurity potential is negative above the Bi atom right over the Zn impurity site.

For the BSCCO surface doped with a Ni impurity, the effective impurity potential also oscillates but changes in the opposite way relative to a Zn impurity. The effective impurity potential is negative in the nearest vicinity of the Ni impurity in a spherical region and becomes positive in the outer cylindrical region as seen in Figure 5(c). Over the Ni impurity and above the Bi atom, the effective impurity potential is positive as shown in Figure 5(d). For the BSCCO surface doped with a vacancy impurity, the effective impurity potential behaves quite differently from those for the Zn and Ni impurities. Since the magnitude of the effective impurity potential is so large for the vacancy impurity, the positive $\Delta V_{eff}(\mathbf{r})$, as seen in Figure 5(e), spans a much larger region than that of the Zn impurity. The oscillation of the effective impurity potential occurs only very far from the vacancy impurity site as shown in Figure 5(f). As a result, the Bi atom right over the vacancy impurity site is still in the range of positive $\Delta V_{eff}(\mathbf{r})$.

To get a quantitative picture of the change of effective potential induced by the impurities and further modified by the interference among the BiO and $CuO_2$ wave functions, $\Delta V_{eff}(\mathbf{r})$ is plotted for different high-symmetry directions passing through the impurity site in Figure 6. These plots show that the Zn, Ni and vacancy impurity induce +9.6 eV, −3.7 eV and +60 eV change in effective potential in the center of the impurity, respectively, with an effective range of about 1 Å. Although as discussed below, the effective potential tails at the surface are most important for the observed STM patterns, it is the strong effective potential within this 1 Å which determines the dynamics of the electron in the $CuO_2$ plane and thus the resonant frequencies in the superconducting state. Small Friedel-type oscillations due to the screening effect from the electrons around the impurities begin at around this distance. Alone among the three impurities considered, the vacancy does not show any sign of oscillation within 2 Å around the impurity. This is because the electron density is so small in the vicinity of the vacancy that screening is negligible. In the inset of Figure 6(a), the change of effective potential is shown around the surface (located at 4.5 Å). For the Zn and Ni impurity, $\Delta V_{eff}(\mathbf{r})$ continues to oscillate and equals about −0.15 V and 0.5 eV at the position of 1.5 Å above the surface. This position corresponds to the largest change of $\Delta V_{eff}(\mathbf{r})$ for Zn and Ni impurities above the surface. For the vacancy, $\Delta V_{eff}(\mathbf{r})$ only starts to oscillate around the surface and has about the same magnitude as that for the Zn and Ni impurities.

According to our calculations, there are some remarkable similarities observed among the three effective impurity potentials. First of all, they are all short range, with range of order 1 Å, and display a much weaker long range tail, which is important for residual impurity effects such as the STM images, but not for the existence or energy of the resonant state in the superconductor. This is in contrast to some theories of the STM impurity spatial pattern, which assume a potential which changes sign but has significant amplitude over the distance of several lattice constants[13]. This does not rule out the viability of such explanations directly, but suggests that the extended nature, if correct, might be due to strong correlation effects neglected in the present calculations. Indeed, it is known from NMR studies that magnetization oscillations induced by the impurity extend over several lattice spacing, but it is far from clear that the effective impurity potential averaged over spin is significantly modified. The short-range character of the potential found in these calculations lends support to those theoretical treatments which treat planar impurities in the cuprates as point-like in character, and suggests that the main difference among the three, at least as far as the CuO$_2$ plane is concerned, is simply their overall strength.

It is tempting to test this hypothesis by comparing with resonance energies determined by STM in the superconducting state, which can be done by calculating the matrix element $V_0$ which is the bare potential entering phenomenological theories of the impurity resonance in the superconducting state. These models[7] consist of a single tight-binding band of electrons $\varepsilon_\mathbf{k}$ near the Fermi level, with single on-site impurity potential $V_0$. We have not attempted to derive such a model formally from the full Kohn-Shan equations to derive the correct effective $V_0$. Instead, we simply assume the DFT effective impurity potential integrated over a unit cell represents a crude approximation to the true $V_0$. (This procedure needs to be justified by a microscopic calculation, but we present the results which obtain from it anyway because they are intriguing; they are not central to our conclusions.) With this assumption, the correct generalization of the resonance frequencies given by Eq. (1) for pointlike potentials to an arbitrary single band are the zeros of the $T$-matrix determinant $(V_0^{-1} - G_3)^2 - G_0^2$, where $G_0 = \Sigma_{\mathbf{k}} \omega / (\omega^2 - \varepsilon_{\mathbf{k}}^2 - \Delta_{\mathbf{k}}^2)$ and $G_3 = \Sigma_{\mathbf{k}} \varepsilon_{\mathbf{k}} / (\omega^2 - \varepsilon_{\mathbf{k}}^2 - \Delta_{\mathbf{k}}^2)$, and $\varepsilon_{\mathbf{k}}$ is the electronic dispersion measured relative to the Fermi level. Note that Eq. (1) follows from these definitions in the limit of perfect particle hole symmetry, $G_3 = 0$. The dependence of the resonance energy $\Omega_0$ on $V_0$ within this phenomenological model, taking $\varepsilon_{\mathbf{k}}$ from fits to angle-resolved photoemission data[22] and assuming a $d_{x^2-y^2}$ order parameter

$$\Delta_k = \frac{\Delta_0}{2}\left(\cos k_x a - \cos k_y a\right),$$ with $\Delta_0 = 30$ meV, is shown in Figure 7. Now assuming a Gaussian form for our ab initio $\Delta V_{\mathit{eff}}(\mathbf{r})$, a range of 1 Å, and a lattice constant of 3.5 Å, we estimate matrix elements $V_0$ of 0.78, −0.30, and 4.8 eV-$a^2$ respectively (5.2, -2.0, and 3.2 $t$-$a^2$, taking $t=150$meV[22]) as input to the phenomenological theory for Zn, Ni, and vacancy, respectively. We then find resonance energies of $|\Omega_0| = 1.8$, 10.5, and 0.9 meV for Zn, Ni, and vacancy, respectively, as compared to experimental values of 1.5, 9, and 0.5 meV[4-6]. This is remarkable agreement given the ab initio nature of the calculation of the effective impurity potentials and the nonlinear relation between resonance energy and potential (different from Eq. (1), which is valid only for an idealized particle-hole symmetric band)[36]. We emphasize that the correct prediction of the absolute value of a

resonance energy for a single impurity species, which depends sensitively on some quantities like $\Delta_0$ not known with high precision, is not as significant as the ability to predict the correct relative sizes of the resonance energies for all three impurities for a given phenomenological parameter set.

Before leaving this section, we comment on one point which may confuse the reader. It is frequently assumed in the phenomenological literature that the effective Zn potential is negative, i.e. *attractive* for electrons (see, e.g. Ref. [7]). The basis for this argument is an atomic one: since both Zn and Cu are expected to be in the ++ state in the $CuO_2$ planes, the energies of the highest occupied orbitals would be given by the third ionization energies, −36.3 eV for Cu and −39.7 eV for Zn. Thus one might expect the Zn level to lie 2-3 eV below the relevant Cu levels before hybridization, and anticipate a transfer of electronic charge from Cu to Zn, corresponding to an impurity potential $V_0<0$. The extent to which charge is transferred from $Cu^{++}$ to $Zn^{++}$ depends even at the atomic level on electron affinity as well as ionization potential, however, and the order of the second ionization energies is reversed for the two elements. Thus it is not a trivial matter to predict how charge will transfer in a solid state situation. DFT calculations of CuZn alloys[37,38] and cluster calculations for Zn in YBCO[19] find in fact that the net charge transfer is away from Zn to Cu, so that the effective Zn potential is indeed repulsive for electrons, in agreement with the current calculation.

## D. Simulated STM images and LDOS in $CuO_2$ plane

We now turn to the question of what an STM experiment on the surface of this material should see. For purposes of this paper, we define a "simulated STM image" as the LDOS at a tip height 1.5 Å above the BiO surface in an energy window around the Fermi level, $\rho(\mathbf{r}, e\Delta V)$ measured at this distance, neglecting tip matrix elements. Actual experiments on cuprate surfaces are usually performed at a greater tip height, of 3.5-4.0 Å, but we argue below that the differences in the qualitative features of the STM patterns are insignificant. Images at −50mV, a bias roughly corresponding to superconducting energies in the system, are shown for BSCCO surfaces doped with Zn, Ni and vacancy impurities, and we have verified that the images do not display any strong dependence on bias at this scale, as is to be expected in the normal state of the material (experiments at low $T$ exhibit a strong bias dependence on this scale, but this is due to pair correlations below $T_c$ ). In Figure 8, each panel shows four adjacent unit cells, with the impurity at the center, corners and the half-way point of each side, i.e. periodically repeated to ensure the applicability of the DFT calculation. For comparison to the simulated STM images, the LDOS in the $CuO_2$ plane two layers below is also shown. In Figure 8 (a-b), the positions of the atoms in the respective layers are indicated. In Figure 8(d), we see that the Zn impurity site has the lowest intensity on the plane cutting through the impurity itself, and that the intensity on the nearest neighbor sites is enhanced. This is understandable because the Zn atom has filled $3d$ shells at a energy far below the Fermi level (see Figure 3b). Under a bias of -50 mV, these states do not contribute to the LDOS. One might therefore expect -- based on the usual interpretation of STM images-- that the position directly over the Zn impurity sites and 1.5 Å above the surface would also represent a local minimum in intensity in that plane. However, in the simulated STM image in Figure 8(c), the surface position above the Zn impurity sites actually correspond to the highest intensity, as also seen in experiment[11], see Figure 1 (a-b). This maximum is unexpected if one tries to associate the STM images directly with the LDOS of impurity states in the $CuO_2$ plane; in the context of the present calculation, however, it is clear that for Zn there is weak band mixing between the $CuO_2$ layer and the BiO layer, as shown in the PDOS discussed in section B.

We warn the reader again that we are computing local densities of states for a system in the normal metallic state, whereas nearly all experiments on impurity states are performed in the superconducting state at low temperatures. With such plots we therefore expect to reproduce neither the strong resonant nature of the impurity state, which is caused by the reduction of states near the Fermi level in the d-wave superconductor, nor all the details of the spatial pattern of the LDOS. We believe nevertheless that these results are relevant for the experiments in the superconducting state, for the simple reason that energy differences involved are, in most cases, comparable to or larger than superconducting state energies. In this case, if the impurity potential gives rise to an LDOS pattern at the surface with maxima displaced from their positions in the $CuO_2$ plane in the normal state, this displacement and other gross features are likely to be retained in the superconducting state as well, and need to be accounted for in any complete theory.

The attribution of the maximum at the center of the STM pattern to an effect of hybridization of impurity wavefunctions with the states between the $CuO_2$ and the surface bears some similarity to an explanation which has been discussed previously in the framework of phenomenological tight binding models[14, 15]. In this approach, the presence of the BiO and SrO layers serve to act as a "fork" or "filter" which forces the electron to tunnel with equal probability from the Bi site above the impurity to the Cu sites neighboring the impurity; thus the measured LDOS at the surface is found to reflect a linear combination of the impurity-induced wavefunctions on the nearest-neighbor Cu sites in the $CuO_2$ layer below. In this picture, the "filter" is a characteristic of the BiO and SrO layer, and is identical regardless of which impurity is present. In contrast, we now argue that it is the details of the impurity states and longer-range part of the effective impurity potential, quite different for each of the three cases considered, which play a crucial role.

We next consider the Cu vacancy, which bears a number of similarities to Zn, in that it clearly expels Fermi level states from its location in the $CuO_2$ plane, and thus creates a strong repulsive potential for electrons locally. This property is reflected in Figure 8(f) as a strong intensity minimum at the center of the pattern. On the other hand, the simulated STM image in Figure 8(e) shows, not an intensity maximum at the center, as in the Zn case, but a local minimum surrounded by four more intense maxima. This result is in agreement with experiment[6], and can be traced back to the fact that the vacancy potential just above the BiO layer is still weakly repulsive. This is most likely a consequence of the weaker screening of the vacancy potential, as argued above.

The situation with the Ni impurity, as shown in Figure 9, is at first glance less amenable to a simple interpretation. At the Ni impurity site itself (Figure 9(b)) there is little apparent difference in intensity or symmetry from the surrounding $d_{x^2-y^2}$ Cu sites, although it is contracted to a somewhat smaller area. On the other hand, there is a dramatic increase of intensity on the Ni site when one goes slightly away from the $CuO_2$ plane, which then falls off at larger distances, as illustrated in Figure 9 (c) and (e). Unlike the Zn impurity state, which clearly also has $d_{x^2-y^2}$ symmetry, the Ni impurity state appears to have a lobe extending in the $z$ direction, of $d_{3z^2-r^2}$ symmetry. Note that the spatial pattern of the LDOS at a tip distance of 3.0 Å from the surface is not qualitatively different from the LDOS at 1.5 Å, although the resolution is substantially decreased. There is no similar strong variation of the impurity site LDOS for either the Zn or the Cu vacancy case, and the images at 3.0 Å are similar to those at 1.5 Å as well. The results of Figure 9 are not surprising if we recall that the hybridization of the Ni impurity state with orbitals in the SrO and BiO layers is considerably stronger than for the other impurities. The LDOS at a position 1.5 Å above the surface corresponding to the Ni site (Figure 9 (a) ) exhibits a strong maximum, in agreement with the strongest Ni resonance observed, at positive bias, but in disagreement with the image of the weaker resonance, at negative bias, as shown in Figure 1 (c).  There is furthermore a strong spatial variation as one approaches the BiO layer, with the central maximum disappearing on the layer itself, implying that the impurity is forcing occupation of the Bi $p_z$ orbital.   Note that the effective impurity potential is actually positive over the entire local region (Figure 6(a)),  yet the central sites in patterns 9(a) and (c) are strong maxima.  This suggests that the relatively strong hybridization in the z direction is lowering the energy of a state, which samples the attractive part of the potential several Angstrom away.  Finally, it may be important that the size of the Ni effective potential at the surface layer is comparable to superconducting state energies, implying that pair correlations can strongly alter the patterns, as is indeed observed in the difference between the negative and positive bias resonances.   An explanation for the rotation of the STM pattern for Ni has been claimed by the "filter" theories of Ni[15], as well as by an extended-potential model[13],  and are therefore also expected emerge in the context of our current microscopic approach were we able to include pair correlations.      We will consider these effects in a later publication.

It is furthermore interesting to compare the details of the impurity patterns away from the central site with experiment, but increasingly problematic to do so the further one moves away from the impurity.  This is because the impurity is periodically repeated in our "sample", and interference effects can arise.  Some aspects of the calculated patterns resemble those measured in the superconducting state.   For example, in Figure 8c, we see that the second highest peaks in the STM intensity pattern are indeed achieved on the *next* nearest neighbor sites, in agreement with experiment on Zn[11].  It is not clear how seriously to take this result, however, since these sites are in fact located at the midpoint between two periodically repeated impurities. In general, we feel that detailed aspects of the normal state patterns calculated here should not be compared with experiments in the superconducting state because of these difficulties.

## Conclusions

In summary, we have reported on density functional calculations of impurity states near surfaces in cuprates. In particular, we have examined the local density of states around Zn, Ni and Cu vacancies both at the sample surface and in the $CuO_2$ plane, and argued that none of the current explanations for the spatial form of the impurity resonance observed by STM in experiments on Zn, Ni and vacancy impurities near the BSCCO-2212 surface[5, 11] is likely to be correct. Instead, we have shown a significant band mixing between the states from the $CuO_2$ layer and those from the BiO layer around the Fermi level. While our proposal is thus similar in some respects to "filtering" of the tunneling current by the BiO and SrO layers, it is nonuniversal in the sense that it depends sensitively on the wave-functions of the particular impurity in question. The effective impurity potential derived from these states is found in general to oscillate as the distance from the impurity increases. For Zn, this causes the sign reversal of the effective impurity potential on the impurity site relative to the point 1.5 Å above the surface, which explains why the intensity of the STM patterns is also reversed relative to what is expected in the $CuO_2$ plane itself. The Cu vacancy potential remains positive but is quite weak at the surface, consistent with lower intensity observed in experiment. Some details of the surrounding spatial patterns observed in STM can also be understood through these calculations, but this is less clear. Unlike Zn and the Cu vacancy, Ni is found to induce impurity states near the Fermi level, which hybridize strongly with O in the SrO layer and Bi in the BiO layer. This gives rise to occupation of states of quite different symmetry both in the $CuO_2$ plane and near the BiO surface.

Because the ranges of the strongest parts of the effective impurity potentials are found to be extremely short, of order 1 Å , our calculations lend support to theoretical models of the effective impurity potential which neglect local correlations and treat impurities as point-like potential scatterers for purposes of calculating dynamical properties of in-plane quasi-particles. In this picture, we believe impurity-induced magnetization effects can still be incorporated as the response of the correlated electron gas to the effective impurity potential we have calculated here, despite the fact that correlations themselves were neglected in this calculation. This follows from the fact that the effective potentials are determined in general by states far from the Fermi level. The close agreement of the relative and absolute sizes of the superconducting state resonance energies estimated from the effective potentials obtained from our *ab initio* calculations lends support to the consistency of this picture.

Our work should inject a cautionary note into the discussion of what can be gleaned from the analysis of STM experiments on cuprate surfaces. According to our current understanding, one is measuring neither the LDOS in the $CuO_2$ plane directly, as is often assumed, nor a "filtered" version of these states via a universal tunneling matrix element. This realization should be important for discussion of nano-scale inhomogeneities and charge ordering in the cuprates, as well as the impurity puzzles discussed here. On a more optimistic note, we have argued that modeling impurity states is possible within DFT, so the realization of the original hope of extracting information on the superconducting state from impurity imaging may not be so distant. Extensions of these calculations to study larger samples, as well to include pair correlations directly, are underway.

# Acknowledgement

We are very grateful for helpful discussions with and suggestions from W. A. Atkinson, A.V. Balatsky, J. C. Davis and J.-X. Zhu. We thank L.-Y. Zhu for providing the calculation on resonant energies. Partial support for research was provided by ONR N00014-04-0060 (PJH), DOE/BES DE-FG02-97ER45660 & DE-FG02-02ER45995, by the NSF/ITR program and Oak Ridge National Laboratory Supercomputer Center (HPC and LLW).

**Figure captions:**

Figure 1. Experimental STM impurity spatial patterns at resonance and associated spectra. (a) observed STM intensity around Zn impurity at $-1.2$mV; (b) comparison of spectra directly above Zn impurity site (red) and at site far from impurity (blue) ; (c) observed STM intensity around Ni impurity site at +9mV (inset: pattern at $-9$mV); (d) comparison of spectra directly above Ni impurity site (red) and away from Ni (blue); (e) observed STM intensity around native defect, possibly Cu vacancy at $-0.5$ mV; (f) comparison of spectra directly above native defect (thick solid line) and away from defect (thin solid line). Data from references 4-6; reproduced with permission of authors.

Figure 2. Structure of bulk BSCCO and the relaxed structure of impurity doped BSCCO surfaces. Panel (a) depicts the pseudo-tetrahedral substructure unit cell of bulk BSCCO. Panel (b), (c) and (d) depict the relaxed structure of the BSCCO surfaces doped with Zn, Ni and vacancy impurity, respectively. Each surface unit cell is $\left(2\sqrt{2}\times2\sqrt{2}\right)R45°$ of the tetrahedral unit cell in panel (a) and doped with one impurity. The arrows on the atoms and the numbers show the directions and magnitudes of the atoms move with respect to their positions in the clean surface. For a Zn impurity, as shown in panel (b), it moves upward by 0.20 Å, the Bi atom above it moves downward by 0.08 Å and the Cu atom below it moves downward by 0.01 Å. For a Ni impurity, as shown in panel (c), it moves downward by 0.15 Å, the Bi atom above it moves upward by 0.03 Å and the Cu atom below it moves upward by 0.07 Å. For a vacancy impurity, as shown in panel (d), the Bi atom above it moves upward by 0.03 Å and the Cu atom below it moves downward by 0.09 Å.

Figure 3. The total density of states and the partial density of states projected on each atomic species. Panel (a), (b), (c) and (d) are for the clean BSCCO surface and the BSCCO surfaces doped with Zn,

Ni and vacancy impurity, respectively.

Figure 4. The partial density of states projected on individual atom. Panel (a) and (b) depict the partial density of states projected on the 3$d$ orbital of the Cu atom below the impurity and that projected on the 6$p$ orbital of the Bi atom above the impurity, respectively. In both panels, the solid, dashed, dotted and dashed-dotted lines are for the clean BSCCO surface and the BSCCO surfaces doped with Zn, Ni and vacancy impurity, respectively.

Figure 5. The iso-surfaces of the difference of effective potential, $\Delta V_{eff}(\mathbf{r})$. Panel (a)-(b), (c)-(d), and (e)-(f) are for the BSCCO surfaces doped with Zn, Ni and vacancy impurity, respectively. The difference is taken between the electronic effective potential of the doped surface and that of the clean surface. In each panel, the darker (red) and lighter (yellow) region correspond to negative and positive regions of the electronic effective potential. The iso-surface values are ±0.5 V in (a), ±0.05 V in (b), ±0.1 V in (c), ±0.05 V in (d), ±1.0 V in (e) and ±0.12 V in (f).

Figure 6. The difference of effective potential, along high symmetry directions passing through the impurity: (a) 001; (b) 100; (c) 110 versus displacement in Angstrom. The solid, dashed and dotted lines are for Zn, Ni and vacancy impurities, respectively. The impurity is placed at the origin (0,0,0). The inset shows the details of $\Delta V_{eff}(\mathbf{r})$ around the surface. The 1$^{st}$ nearest neighboring Cu site is 3.67 Å away from the impurity site, which is indicated by the vertical dashed line in the inset of panel (c).

Figure 7. Resonance energy for impurity bound state within phenomenological one-band model

from Ref. [22], and $\Delta_0$=30meV, with impurity matrix elements $V_0$ estimated from ab initio calculations. Energies for three impurities indicated by arrows.

Figure 8. LDOS $\rho(\mathbf{r},\omega)$ for Zn and Cu vacancy at $\omega$= −50 meV. All left panels correspond to simulated STM images defined as LDOS on a plane at distance 1.5 Å above the BiO surface, all right panels to the $CuO_2$ plane. Panels (a)-(b), schematic picture of atoms in BiO and $CuO_2$ layers; (c)-(d) LDOS with Zn impurity replacing Cu in $CuO_2$; (e)-(f) similar with Cu vacancy.

Figure 9. LDOS $\rho(\mathbf{r},\omega)$ for Ni impurity at $\omega$= −50 meV. a) , c), e) LDOS above the BiO surface, at distance 3 Å, 1.5 Å, and 0.3 Å, respectively; b), d), f) LDOS near the $CuO_2$ plane, 0 Å, 0.3 Å, and 0.6 Å below the plane.

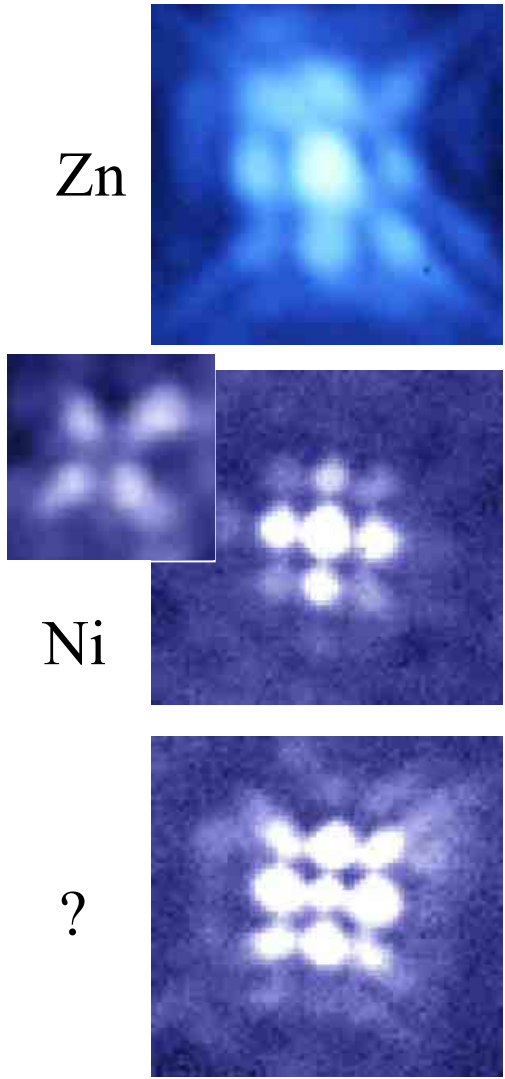
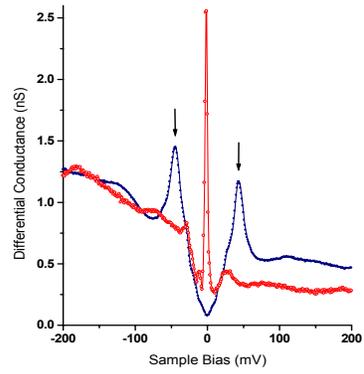
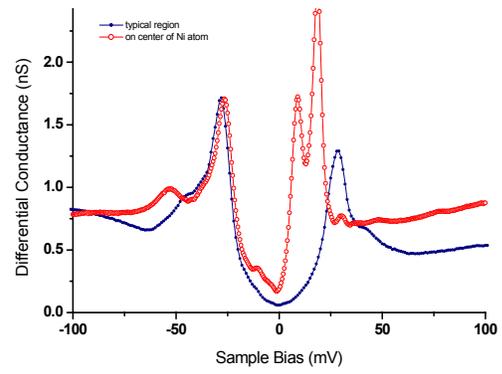
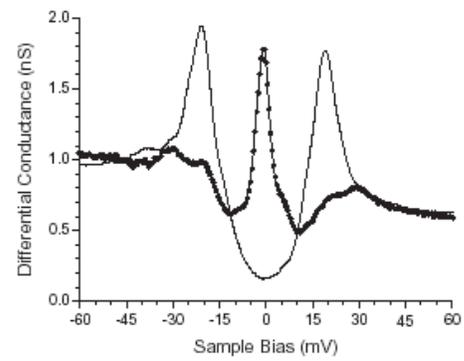

Figure 1

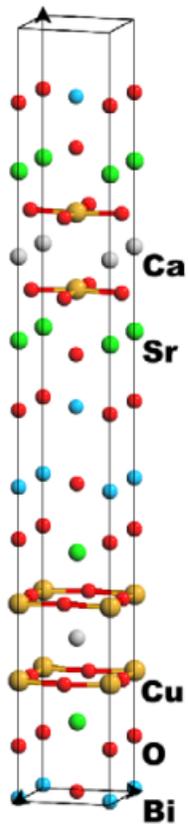

(a)

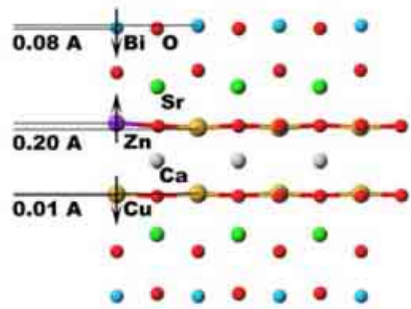

(b)

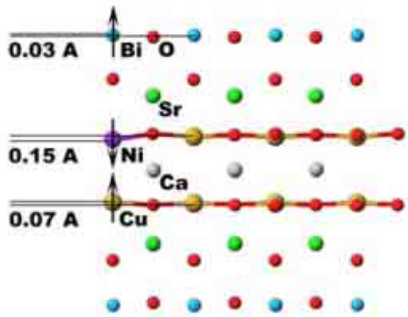

(c)

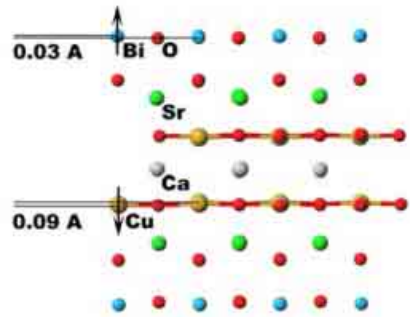

(d)

Figure 2

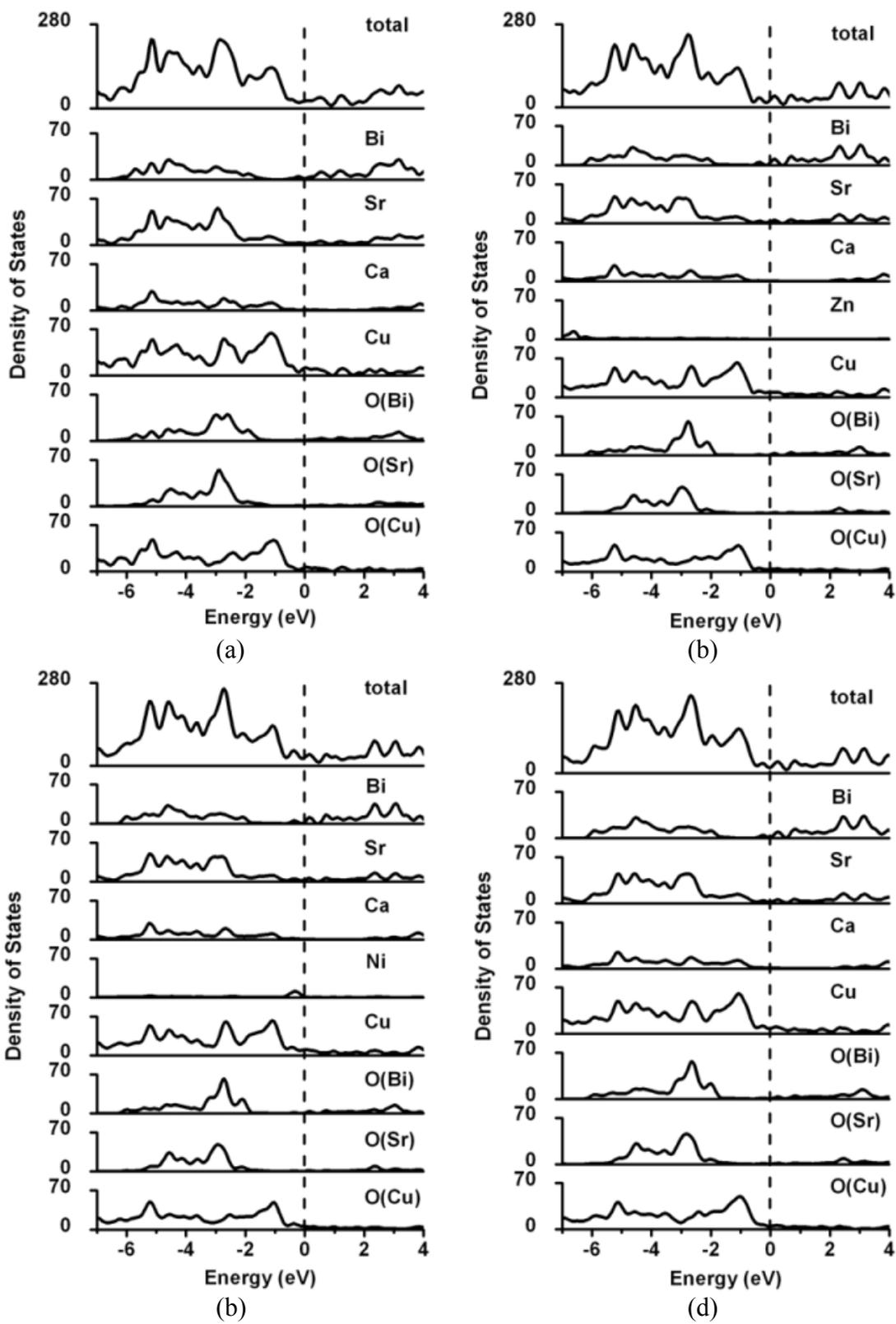

Figure 3

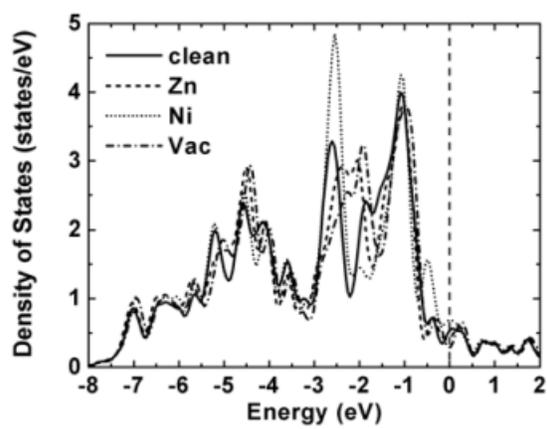 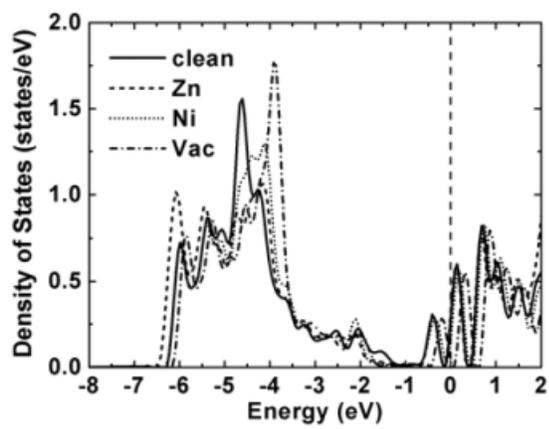

(a) (b)

Figure 4

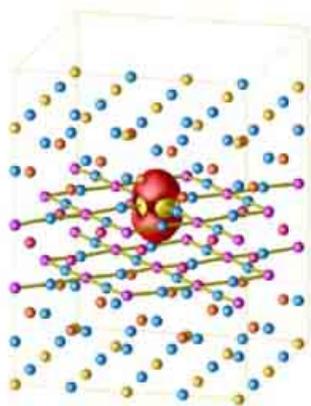
(a)

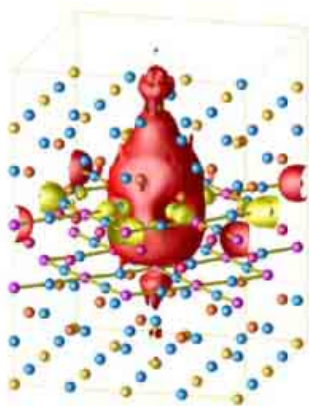
(b)

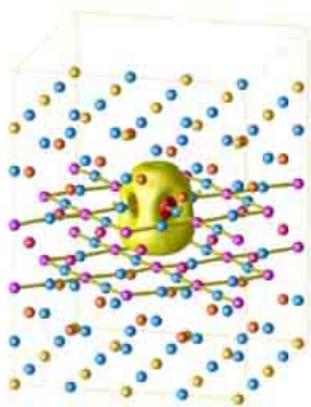
(c)

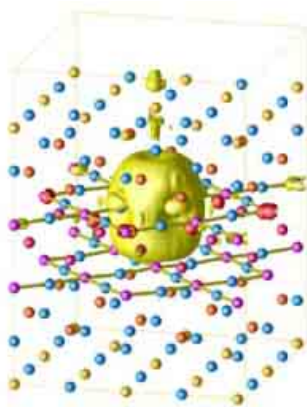
(d)

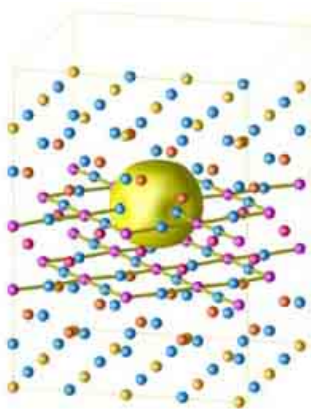
(e)

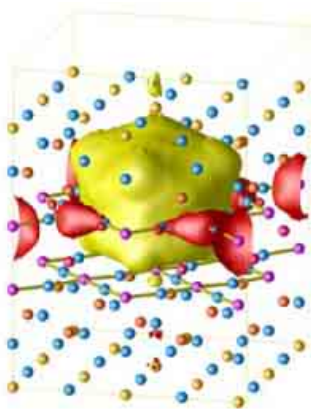
(f)

Figure 5

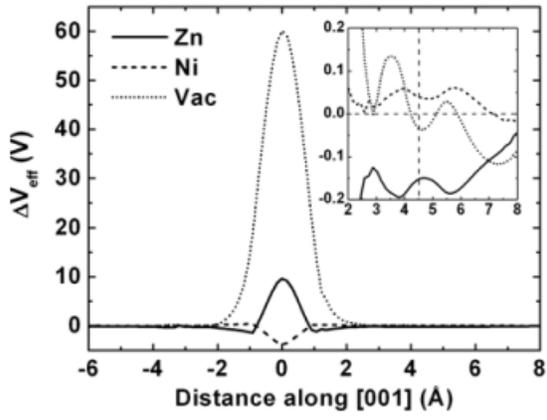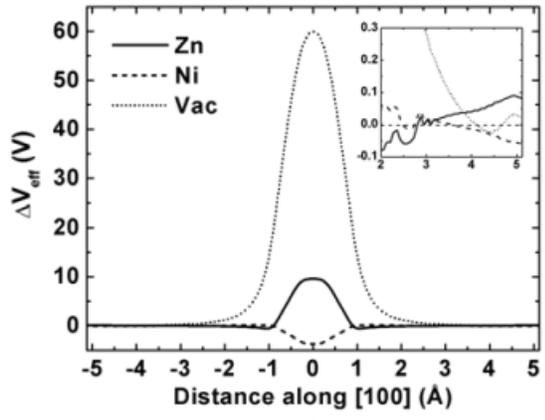

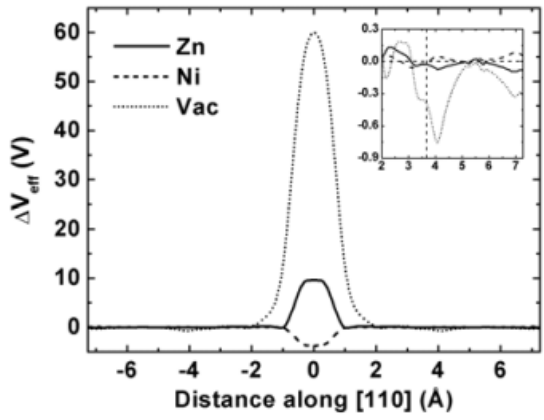

Figure 6

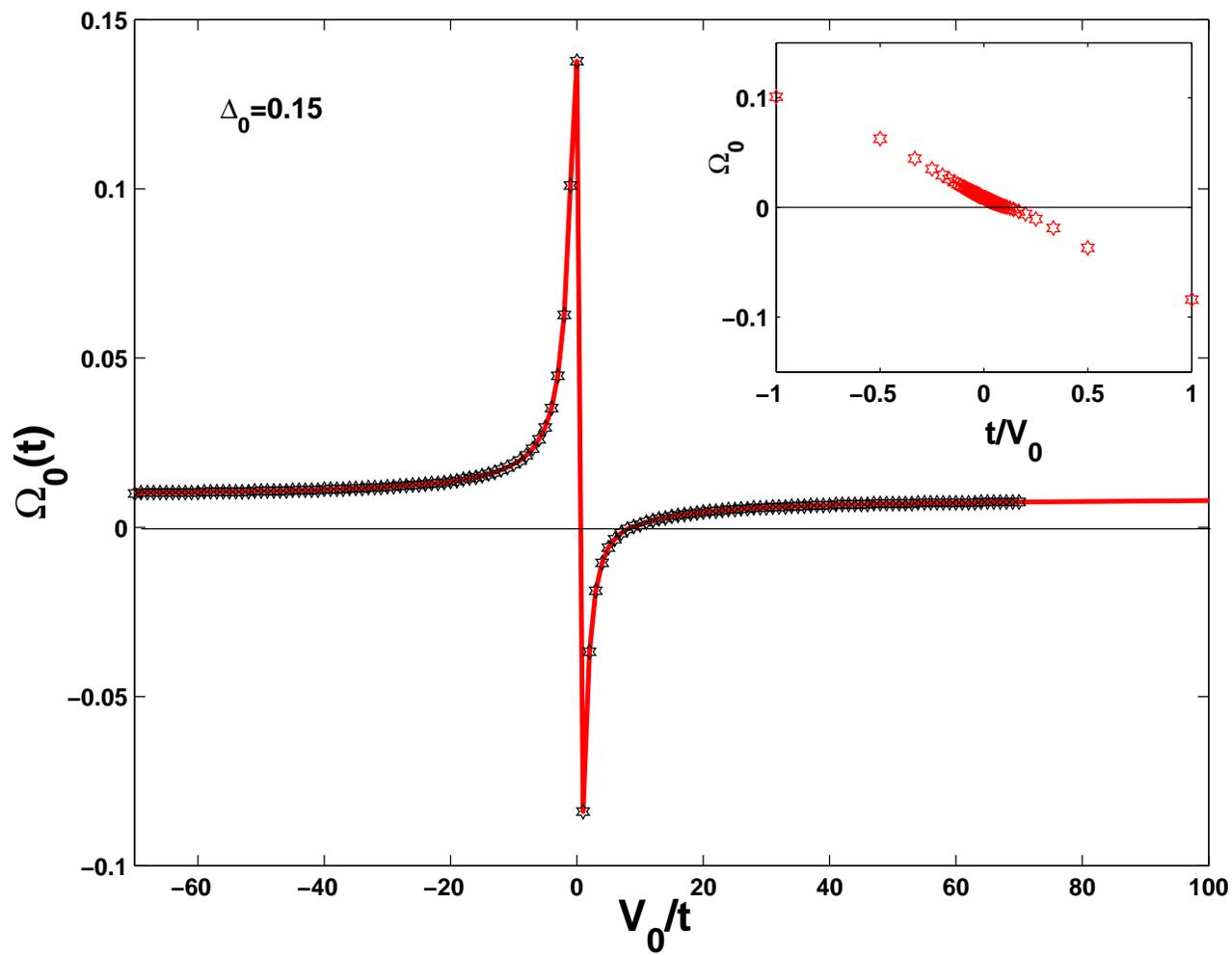

Figure 7

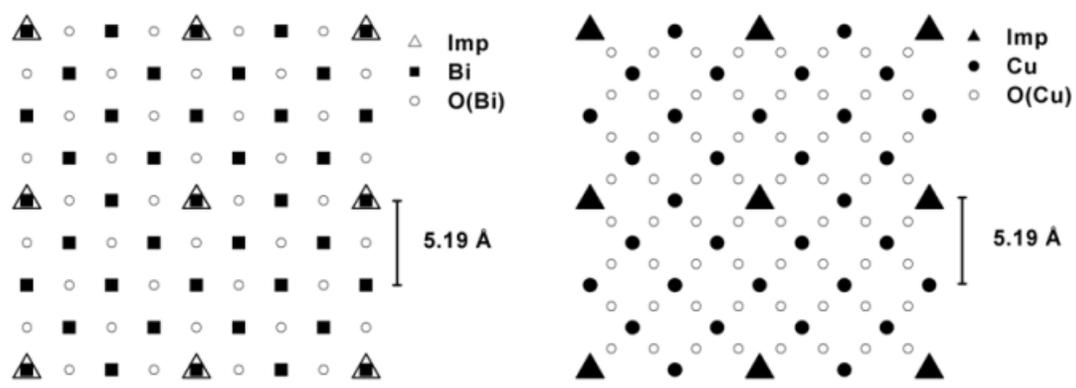

(a)  (b)

Zn 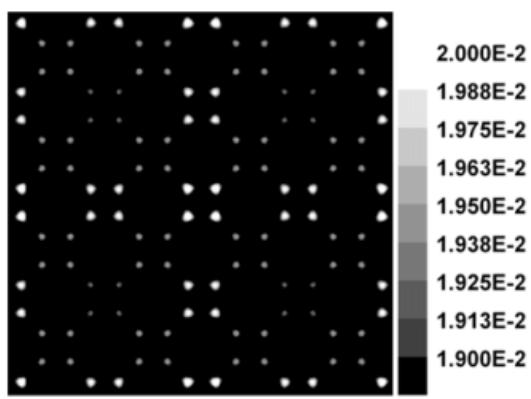 (c)  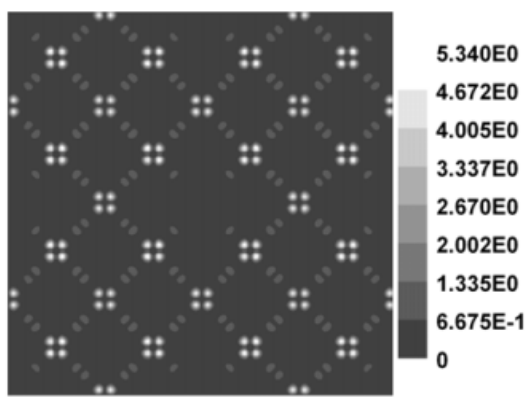 (d)

Vacancy 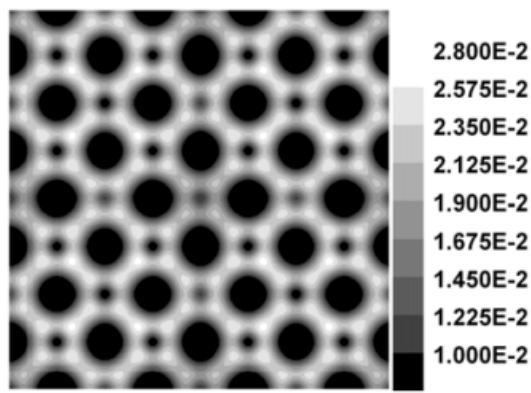 (e)  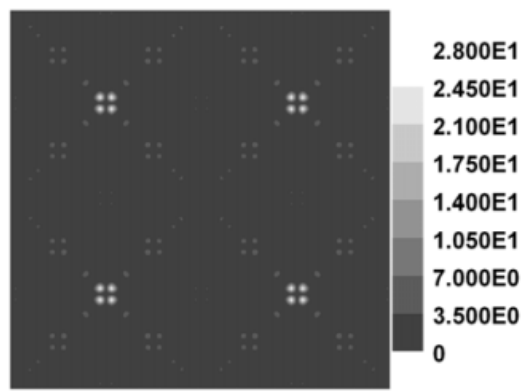 (f)

Figure 8

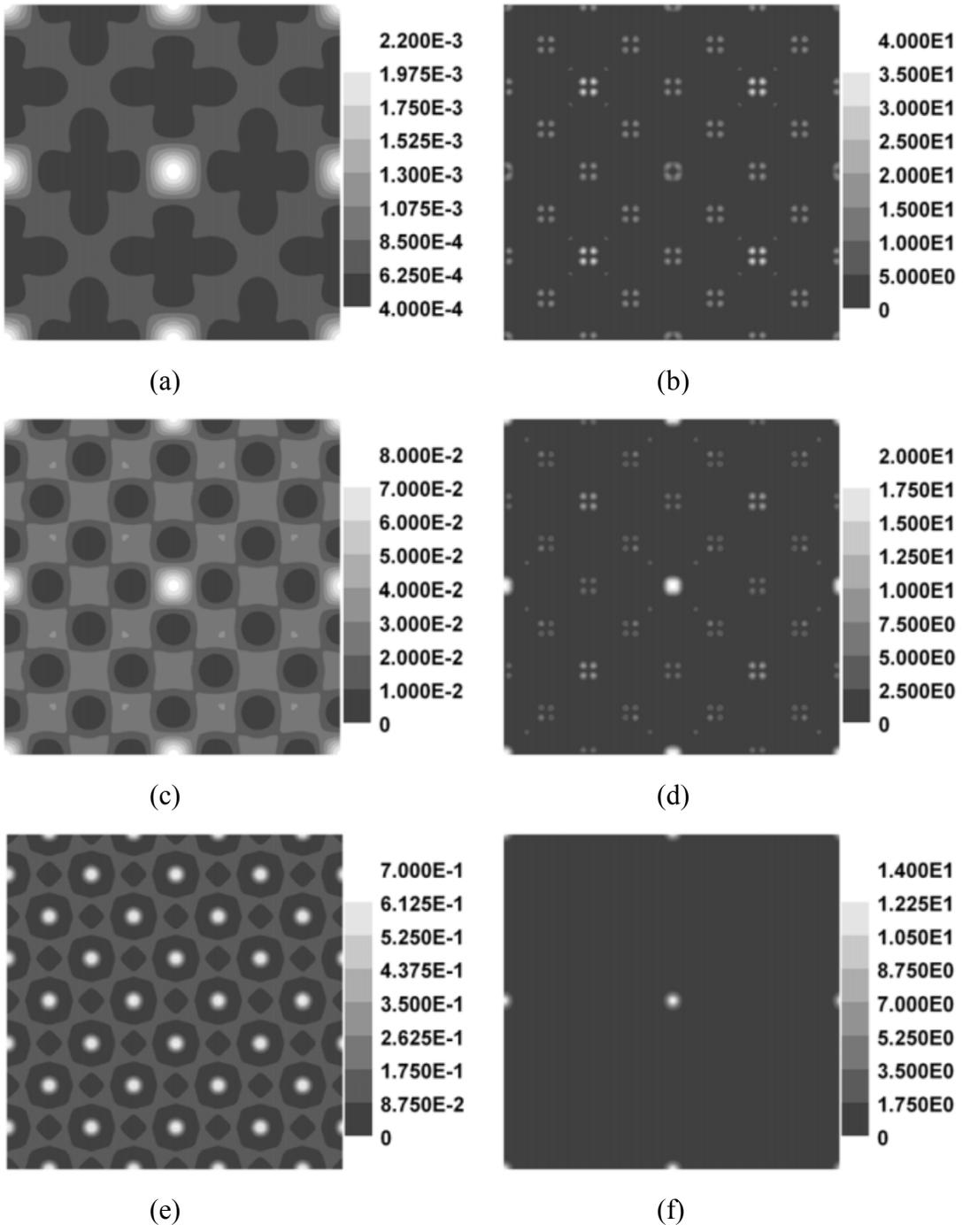

Figure 9